\documentstyle[prl,aps,epsf]{revtex}

\begin{document}

\draft

\title{Divergence-free WKB method}

\author{Tadanori Hyouguchi${}^1$, Satoshi Adachi${}^1$,  
and Masahito Ueda${}^{1,2}$}

\address{
${}^1$Department of Physics, Tokyo Institute of Technology, Meguro-ku,
Tokyo 152-8551, Japan \\
${}^2$CREST, Japan Science and Technology Corporation (JST), Saitama
332-0012, Japan}

\date{\today}

\maketitle

%%%%%%%%%%%%%%{abstract}%%%%%%%%%%%%%%%%%%%%%%%%%%%%%%%%%%%%%%%%%%%%%%%%%

\begin{abstract}
A new semiclassical approach to linear (L) and nonlinear (NL)
one-dimensional Schr\"odinger equation (SE) is presented. 
Unlike the usual WKB solution, our solution does not diverge 
at the classical turning point.
For LSE, our zeroth-order solution, when expanded in powers of $\hbar$, 
agrees with the usual WKB solution ${e^{i S/{\hbar}}}/{\sqrt{v}}$ up to 
${\cal O}(\hbar^0)$, where $S$ and $v$ are the classical action and the
velocity, respectively.
For NLSE, our zeroth-order solution includes quantum corrections to
the Thomas-Fermi solution, 
thereby giving a smoothly decaying wave function into the forbidden region. 
\end{abstract}

\pacs{03.65.Ge, 03.65.Sq}

%\narrowtext

The WKB method allows us to derive expressions 
for various quantum-mechanical quantities 
when the action is much larger than $\hbar$,  
and has been widely used in many subfields of physics and
chemistry~\cite{Ber,Fro,Mil,Vor}.
Nevertheless, the WKB method has a serious flaw of divergence 
at the classical turning point, 
which originates from the fact that the classical trajectry
is taken as the zeroth-order solution.  
Considerable efforts have been devoted to overcome this problem, 
among other things, the complex method~\cite{Ber,Fro} 
and the uniform approximation~\cite{Lan,Goo}.
However, the former method cannot describe the wave function at the
turning point, while the latter one requires us to resort to a  
comparison function \cite{Din} that mimics the original potential 
and to which an analytic solution is available. 
This Letter presents a new semiclassical method 
whose zeroth-order solution is constructed upon a trajectry 
with quantum corrections such that the 
aforementioned divergence is absent. 
Moreover, our method gives a solution to the nonlinear Schr\"odinger
equation (NLSE) on an equal footing.

%%%%%%%%%%%%%%{Section.1}%%%%%%%%%%%%%%%%%%%%%%%%%%%%%%%%%%%%%

We begin by reviewing the WKB method for the one-dimensional 
linear Schr\"odinger equation (LSE)
$- \frac{\hbar^2}{2m}\frac{d^2 \psi}{d x^2} + V \psi = E \psi$. 
Rescaling the length and the energy in units of $l$ and $\hbar^2/{2ml^2}$, 
respectively,  where $l$ is a characteristic length scale of the
potential $V(x)$,  
LSE takes the form 
\begin{equation}
- \psi'' + V \psi = E \psi ~,
\label{1-2}
\end{equation}
where the prime denotes the differentiation with respect to $x$. 
In Eq.~(\ref{1-2}), the length $x$ and the energies $E,V$ are 
proportional to $\hbar^0$ and $\hbar^{-2}$, respectively.  
We introduce the action function $\varphi(x)$ through
$\psi(x) = e^{\varphi(x)}$ ,
where the action is measured in units of $\hbar$. 
In terms of $\varphi(x)$, Eq.~(\ref{1-2}) takes the form 
\begin{equation}
(\varphi')^2 + (E-V) = - \varphi'' ~.
\label{1-4}
\end{equation}
The zeroth-order WKB solution $\varphi'_{\rm WKB,0}$ , which is obtained
by neglecting $\varphi''$ in Eq.~(\ref{1-4}), satisfies the 
following equation:
\begin{equation}
(\varphi'_{\rm WKB,0})^2 + (E-V) = 0 ~.
\label{1-5}
\end{equation}
The classical turning point $x^{\rm (c)}$ is determined by $V(x^{\rm (c)})=E$.
Incorporating the effect of $\varphi''$ perturbatively, 
the WKB solution $\varphi'_{\rm WKB}$ takes familiar forms as
\begin{equation}
\varphi'_{\rm WKB} = \left\{
\begin{array}{ll}
  \pm i & \sqrt{E-V} - \frac{1}{4}\frac{V'}{V-E} + \cdots ~~(V<E) ,\\
  \pm   & \sqrt{V-E} - \frac{1}{4}\frac{V'}{V-E} + \cdots ~~(V>E) .\\
\end{array} \right.
\label{1-7}
\end{equation}
We note that the WKB solution diverges at $x^{\rm (c)}$.
This divergence originates from the fact that 
the classical trajectry is taken as the zeroth-order solution. 
Our strategy is to incorporate quantum corrections 
in  the zeroth-order solution so as to remove the divergence. 

%%%%%%%%%%%%%%%{Section.2}%%%%%%%%%%%%%%%%%%%%%%%%%%%%%%%%%%%%%%%%

To carry out this program,  
we differentiate both sides of Eq.~(\ref{1-4}) with respect to $x$,
having 
\begin{equation}
2 \varphi' \varphi'' - V' = - \varphi''' ~, 
\label{2-1}
\end{equation}
and substitute $\varphi''$ in Eq.~(\ref{1-4}) into Eq.~(\ref{2-1}), 
obtaining
\begin{equation}
(\varphi')^3 + (E-V)\varphi' + \frac{1}{2}V' = \frac{1}{2}\varphi''' ~.
\label{2-2}
\end{equation}
The Schr\"odinger equation~(\ref{1-2}) is sufficient for Eq.~(\ref{2-2})
to hold, but it is not necessary. 
To show this, we add $\varphi' \varphi''$ to both sides of
Eq.~(\ref{2-2}) and integrate the resulting equation 
with respect to $x$. We then get 
\begin{equation}
g e^{2 \varphi} = (\varphi')^2 + (E-V) + \varphi'' ~, 
\label{2-4}
\end{equation}
where $g$ is a constant of integration. Equation (\ref{2-4}) can be rewriten as
\begin{equation}
- \psi'' + V \psi + g \psi^3 = E \psi ~.
\label{2-5}
\end{equation}
This is nothing but NLSE which includes LSE 
as a particular case of $g=0$. 
Throughout this Letter, we assume that for NLSE the wave function 
is normalized to unity
$\int_{-\infty}^{\infty} dx~|\psi(x)|^2 = 1$, 
while for LSE the normalization constant is left arbitrary.

%%%%%%%%%%%%%%%%{Section.3}%%%%%%%%%%%%%%%%%%%%%%%%%%%%%%%%%%%%%%%

Our zeroth-order solution $\varphi'_0$, which is obtained by neglecting
$\varphi'''$ in Eq.~(\ref{2-2}), satisfies the following cubic equation
\begin{equation}
(\varphi'_0)^3 + 3p \varphi'_0 + 2q = 0 ~,
\label{3-1}
\end{equation}
where $p \equiv (E-V)/3$ and $q \equiv  V'/4$.
Comparing Eqs.~(\ref{2-2}) and (\ref{3-1}) 
with Eqs.~(\ref{1-4}) and (\ref{1-5}), respectively, 
we see our method is a natural extension of the WKB method. 
Equation (\ref{3-1}) is the fundamental equation of 
our new semiclassical method.
As we shall see later, the solutions of Eq.~(\ref{3-1}) lead to  
divergence-free wave functions. 

A root of the determinant of Eq.~(\ref{3-1}), $D \equiv p^3+q^2=0$, 
determines our turning point $x=x^{\rm (q)}$, which includes 
quantum corrections. 
In fact, the difference between $x^{\rm (q)}$ and $x^{\rm (c)}$ 
may be expanded as 
\begin{equation}
  x^{\rm (q)} - x^{\rm (c)} 
= \frac{3}{2}\frac{1}{ [ 2 V'(x^{\rm (c)}) ]^{1/3}} 
  \left( 1 + \frac{\epsilon}{2} + {\cal O}(\epsilon^2 , \delta) \right) ~,
\label{3-3}
\end{equation}
where
\[ 
 \epsilon \equiv 
 \frac{V''(x^{\rm (c)})}{ [2V'(x^{\rm (c)})]^{4/3} } ~~{\rm and}~~
 \delta \equiv 
\frac{V'''(x^{\rm (c)})}{ [2V'(x^{\rm (c)})]^{5/3} }~.
\]

%%%%%%%%%%%%%%%%{Section.4}%%%%%%%%%%%%%%%%%%%%%%%%%%%%%%%%%%%%

We first discuss the zeroth-order solution.
In the allowed region, the cubic equation~(\ref{3-1}) has 
the following three solutions:
\begin{equation}
\varphi'_0 = \left\{
\begin{array}{lll}
  - \kappa , \\
  \frac{1}{2}\kappa + ik \equiv \varphi'_+ ~, \\
  \frac{1}{2}\kappa - ik \equiv \varphi'_- ~, \\
\end{array}\right.
\label{4-1}
\end{equation}
where
\begin{equation}
\left\{ 
\begin{array}{lllll}
  \kappa(x) \equiv &  
  & (q+\sqrt{D})^{1/3} + (q-\sqrt{D})^{1/3} & ~,\\
  k(x) \equiv & \frac{\sqrt{3}}{2} 
  & \left[ (q+\sqrt{D})^{1/3} - (q-\sqrt{D})^{1/3} \right] & ~.\\
\end{array}
\right.
\label{4-2}
\end{equation}
When expanded in powers of $\hbar$, Eq.~(\ref{4-1}) reduces to  
\begin{equation}
\varphi'_0 = \left\{
\begin{array}{lllllllllll}
- \kappa   =& &            &-& \frac{2q}{3p} + {\cal O}(\hbar^1)   
           =& &            & & \frac{1}{2}\frac{V'}{V-E} 
           + {\cal O}(\hbar^1) ~,\\
\varphi'_+ =& & i\sqrt{3p} &+& \frac{ q}{3p} + {\cal O}(\hbar^1)
           =& & i\sqrt{E-V}&-& \frac{1}{4}\frac{V'}{V-E}
           + {\cal O}(\hbar^1) ~,\\
\varphi'_- =&-& i\sqrt{3p} &+& \frac{ q}{3p} + {\cal O}(\hbar^1)         
           =&-& i\sqrt{E-V}&-& \frac{1}{4}\frac{V'}{V-E} 
           + {\cal O}(\hbar^1) ~.\\
\end{array}\right.
\label{4-3}
\end{equation}
where $-\kappa$ and $\varphi'_{\pm}$, 
respectively, correspond to the Thomas-Fermi solution~\cite{Gol,Bay}
and the WKB solutions (\ref{1-7}) up to ${\cal O}(\hbar^0)$.
Substituting Eq.~(\ref{4-3}) into Eq.~(\ref{2-4}), we obtain
\begin{equation}
g \psi^2 = \left\{
\begin{array}{lll}
  E - V + & {\cal O}(\hbar^0) ~~{\rm for}~~ \varphi'_0 = - \kappa   ~,\\
          & {\cal O}(\hbar^0) ~~{\rm for}~~ \varphi'_0 = \varphi'_+ ~,\\
          & {\cal O}(\hbar^0) ~~{\rm for}~~ \varphi'_0 = \varphi'_- ~.\\
\end{array}\right. 
\label{4-4}
\end{equation}
From Eq.~(\ref{4-4}), it is clear that $-\kappa$ and $\varphi'_{\pm}$ give 
a solution to NLSE and those to LSE, respectively.

%%%%%%%%%%%%%%%{Section.5}%%%%%%%%%%%%%%%%%%%%%%%%%%%%%%%%%%%%%%%%%%%%%%%%%%%

In the forbidden region, the cubic equation (\ref{3-1}) 
has the following three real solutions :
\begin{equation}
\varphi'_0 = \left\{
\begin{array}{lll}
- \kappa &=& \mp 2\sqrt{-p}\cos
  \left( \frac{1}{3} {\rm Arctan} \frac{\sqrt{-D}}{|q|}                 
  \right) ~,\\
  \chi'_+ &=& \mp 2\sqrt{-p}\cos
  \left( \frac{1}{3} {\rm Arctan} \frac{\sqrt{-D}}{|q|}+\frac{2\pi}{3} 
  \right) ~,\\
  \chi'_- &=& \mp 2\sqrt{-p}\cos
  \left( \frac{1}{3} {\rm Arctan} \frac{\sqrt{-D}}{|q|}-\frac{2\pi}{3} 
  \right) ~,\\
\end{array}\right.
\label{5-1}
\end{equation}
where the $-$ and $+$ signs correspond to $V'>0$ and $V'<0$, 
respectively. 
When expanded in powers of $\hbar$, Eq.~(\ref{5-1}) reduces to 
\begin{equation}
\varphi'_0 = \left\{
\begin{array}{llllllllll}
- \kappa =& \mp \sqrt{-3p}&+& \frac{ q}{3p} + {\cal O}(\hbar^1)
         =& \mp \sqrt{V-E}&-& \frac{1}{4}\frac{V'}{V-E} 
         + {\cal O}(\hbar^1) ~,\\
 \chi'_+ =& \pm \sqrt{-3p}&+& \frac{ q}{3p} + {\cal O}(\hbar^1)         
         =& \pm \sqrt{V-E}&-& \frac{1}{4}\frac{V'}{V-E} 
         + {\cal O}(\hbar^1) ~,\\
 \chi'_- =&               &-& \frac{2q}{3p} + {\cal O}(\hbar^1)   
         =&               & & \frac{1}{2}\frac{V'}{V-E} 
         + {\cal O}(\hbar^1) ~,\\
\end{array}\right.
\label{5-2}
\end{equation}
where $-\kappa$ and $\chi'_+$ correspond to the WKB solutions (\ref{1-7}) 
up to $O(\hbar^0)$.
Substituting Eq.~(\ref{5-2}) 
into Eq.~(\ref{2-4}), we obtain
\begin{equation}
g \psi^2 = \left\{
\begin{array}{llll}
        & & {\cal O}(\hbar^0) ~~{\rm for}~~ \varphi'_0 = - \kappa  ~,\\
        & & {\cal O}(\hbar^0) ~~{\rm for}~~ \varphi'_0 =   \chi'_+ ~,\\
  E - V &+& {\cal O}(\hbar^0) ~~{\rm for}~~ \varphi'_0 =   \chi'_- ~.\\
\end{array}\right.  
\label{5-3}
\end{equation}

%%%%%%%%%{Section.6}%%%%%%%%%%%%%%%%%%%%%%%%%%%%%%%%%%%%%%%%%%%%%

To proceed further with our analysis, we shall assume that 
$V'(x) \ge 0 ~~(\forall x)$.
Then the wave function must decay to zero as $x \to \infty$, {\it i.e.},
$\displaystyle \lim_{x \to \infty} \psi(x)=0$.
Consequently, in the forbidden region, $-\kappa$ must be chosen as the
zeroth-order solution $\varphi'_0$ for both cases of LSE and NLSE. 
Note that, for NLSE,  $g \psi^2 = {\cal O}(\hbar^0)$ in Eq.~(\ref{5-3}) 
does not mean that $g=0$ but that the wave function is well damped 
in the forbidden region.

We are now in a position to construct our zeroth-order wave functions
to LSE and NLSE. 
For NLSE, the zeroth-order wave function is
\begin{equation}
  \psi_0(x) = N \exp \left( - \int_{x^{\rm (q)}}^x dx' ~\kappa(x') \right) ~,
\label{6-2}
\end{equation}
where $N$ is a normalization constant.
Note that this single solution covers both allowed and forbidden regions.

%%%%%%%%%%{Section.7}%%%%%%%%%%%%%%%%%%%%%%%%%%%%%%%%%%%%%%%%%%%%%%%%%%%%%%

For LSE, the zeroth-order wave function can be described, 
in general, as follows
\begin{equation}
\psi_0(x) = \left\{ 
\begin{array}{llll}
  A_+ \exp \left( \int_{x^{\rm (q)}}^x dx'~ \varphi'_+(x') \right)
+ A_- \exp \left( \int_{x^{\rm (q)}}^x dx'~ \varphi'_-(x') \right) 
\equiv \psi_{\rm I}(x) 
& (x < x^{\rm (q)})  ~,\\
  \exp \left(-\int_{x^{\rm (q)}}^x dx'~ \kappa(x') \right) 
\equiv \psi_{\rm II}(x)
& (x > x^{\rm (q)})  ~.\\ 
\end{array}\right.    
\label{7-1}                       
\end{equation}
To determine the relation between constants $A_+$ and $A_-$, 
we note that, near $x^{\rm (c)}$, LSE (\ref{1-2}) reduces to
\begin{equation}
-\frac{d^2 \psi}{d \xi^2} + \xi \psi = 0 ~, 
\label{7-2}
\end{equation}
where $\xi \equiv [ V'(x^{\rm (c)}) ]^{1/3}(x - x^{\rm (c)})$.
The exact solution of Eq.~(\ref{7-2}) 
that satisfies the boundary condition 
$\displaystyle \lim_{\xi \to \infty} \psi(\xi)=0$ 
is the Airy function ${\rm Ai}(\xi)$~\cite{Abr}.
In the region $|\xi| \gg 1$, 
the asymptotic forms of ${\rm Ai}(\xi)$ are
\begin{equation}
{\rm Ai}(\xi) \sim \left\{
\begin{array}{lll}
  \frac{1}{\sqrt{\pi}}  (-\xi)^{-1/4} 
& \sin\left( \frac{2}{3}(-\xi)^{3/2} + \frac{\pi}{4} \right)
&(\xi < 0)~,\\
  \frac{1}{2 \sqrt{\pi}} \xi^{-1/4} 
& \exp\left(-\frac{2}{3} \xi^{3/2} \right)
&(\xi > 0) ~.\\  
\end{array}\right.
\label{7-3}
\end{equation}
On the other hand, for the linear potential, where 
$p = - [ V'(x^{\rm (c)}) ]^{2/3} \xi/3$ and 
$q =     V'(x^{\rm (c)}) /4$,
one can integrate $\varphi'_{\pm}$ and $\kappa$ in Eq.~(\ref{7-1});  
then the asymptotic behavior of Eq.~(\ref{7-1}) for $|\xi| \gg 1$ becomes
\begin{equation}
\psi_0 = 
\left\{ 
\begin{array}{ll}
\psi_{\rm  I} 
\sim (2e)^{-2/3} 2^{1/3} (-\xi)^{-1/4} 
\left[
A_+ \exp \left(-i \left( \frac{2}{3}(-\xi)^{3/2}+\frac{\pi}{4} \right) 
         \right) + 
A_- \exp \left( i \left( \frac{2}{3}(-\xi)^{3/2}+\frac{\pi}{4} \right) 
         \right)  
\right] \\
\hfill (\xi<0) ~,\\
\psi_{\rm II}
\sim \frac{1}{2} (2e)^{5/6} 2^{1/3} \xi^{-1/4} 
     \exp \left(- \frac{2}{3}\xi^{3/2}\right) 
\hfill (\xi>0) ~.\\
\end{array}\right.                           
\label{7-5}
\end{equation}
For Eq.~(\ref{7-5}) to match Eq.~(\ref{7-3}), we must choose ${A_+}/{A_-}=-1$. 
The zeroth-order solution (\ref{7-1}) can thus be described as \\
\clearpage 
\begin{equation}
\psi_0(x)= \left\{ 
\begin{array}{llll}
\psi_{\rm  I}
= A \exp \left( \frac{1}{2}\int_{x^{\rm (q)}}^xdx'~\kappa(x') \right)
    \sin \left( \int_x^{x^{\rm (q)}}dx'~k(x') \right) 
& (x<x_0) ~,\\
\psi_{\rm II}= \exp \left(-\int_{x^{\rm (q)}}^x dx'~\kappa(x')\right) 
& (x>x_0) ~,\\
\end{array}\right.   
\label{7-6}                        
\end{equation}
where $A \equiv \mp 2i A_{\pm}$.
For $\psi_{\rm I}$ and $\psi_{\rm II}$ and their first derivatives 
to be continuous at a certain point $x_0~( < x^{\rm (q)})$, {\it i.e.}, 
$\psi _{\rm I}(x_0)=\psi _{\rm II}(x_0)$ and
$\psi'_{\rm I}(x_0)=\psi'_{\rm II}(x_0)$,  
$x_0$ and $A$ must satisfy~~~
$  
  \tan \left( \int_{x_0}^{x^{\rm (q)}}dx~k(x) \right) 
= \frac{2 k(x_0)}{3 \kappa(x_0)}
$ 
~~~and~~~
$
A = \exp\left(\frac{3}{2}\int_{x_0}^{x^{\rm (q)}}dx~\kappa(x) \right) 
    {\rm cosec}\left( \int_{x_0}^{x^{\rm (q)}}dx~k(x) \right)
$.
These simulutaneous equations can be solved approximately, giving 
$x_0 = x^{\rm (c)} + {\cal O}(\epsilon)$ and 
$A = (2e)^{3/2} + {\cal O}(\epsilon)$.
For the linear potential $(\epsilon = 0)$, 
the asymptotic behavior of $\psi_{\rm I}$ ($\psi_{\rm II}$)
in Eq.~(\ref{7-5}) for $\xi \to - \infty$ ($+\infty$)
with $A_{\rm \pm} = \mp (2e)^{3/2} / {2i}$ 
exactly agrees with that of ${\rm Ai}(\xi)$ in Eq.~(\ref{7-3})
except for an overall factor.

%%%%%%%%%%{Section.8}%%%%%%%%%%%%%%%%%%%%%%%%%%%%%%%%%%%%%%%%%%%%%%%%%%%%%%

Our zeroth-order wave function (\ref{7-6}) vanishes at 
$x^{\rm (q)}$ for any potential,
which gives us a new quantization condition 
(compared with the WKB quantization condition)
\begin{equation}
\left\{
\begin{array}{llll}
   \frac{1}{2\pi}\oint dx~k(x) 
&= \frac{1}{ \pi}\int_{x_{\rm L}^{\rm (q)}}^{x_{\rm R}^{\rm (q)}}dx 
& k(x) = n+1 
\\          
& & ({\rm Our~method}),
\\
  \frac{1}{2\pi}\oint dx~k_{\rm WKB}(x) 
&= \frac{1}{\pi}\int_{x_{\rm L}^{\rm (c)}}^{x_{\rm R}^{\rm (c)}}dx
& k_{\rm WKB}(x) = n+\frac{1}{2}
\\ 
& & ({\rm the~WKB~method}),\\
\end{array}
\right.
\label{8-1}
\end{equation}
where $k(x)$ is given in Eq.~(\ref{4-2}),
$k_{\rm WKB}(x)\equiv \sqrt{E - V(x)}$, 
$n=0,1,2,\cdots$ is the number of nodes of the wave function, 
and the suffixes L and R denote the left- and right-side turning points, 
respectively.
We note that $k(x)$ and $x^{\rm (q)}$ include quantum corretions, 
which accounts for an extra $1/2$ in our quantization condition. 

We have so far ignored the term $\varphi'''/2$ on the right-hand side of
Eq.~(\ref{2-2}). 
The effect of this term may be evaluated perturbatively, giving~\cite{note}
\begin{equation}
  \varphi'_1 = \frac{1}{27}
\left[ - \varphi'_0 (V')^2 
         \frac{ \left( \varphi'_0 - \frac{V''}{2V'} \right)^2 }
              { \left( (\varphi'_0)^2 + p            \right)^4 }
       + (V')^2 \frac{ \varphi'_0 - \frac{V''}{2V'}        }
                     { \left( (\varphi'_0)^2 + p \right)^3 }
       + \frac{3}{2} V'' \frac{ \varphi'_0 - \frac{V'''}{2V''} }
                              { \left( (\varphi'_0)^2 + p \right)^2 }
\right] ~.
\label{8-2}
\end{equation}
The first-order wave function $\psi_1(x)$ can be constructed by using
$\varphi'_0 + \varphi'_1$ instead of $\varphi'_0$. 
 
We finally apply the zeroth-order solution $\psi_0(x)$ and 
the first-order solution $\psi_1(x)$ to a linear potential (Figs.~1 and 2), 
a parabolic potential $V=x^2$ with $E=17$ (Figs.~3 and 4), 
and the Morse potential $V=900[\exp(-2x) - 2\exp(-x)]$ with $E=-(39/2)^2$ 
(Fig.~5).  
We also compare them with the exact solution and the WKB solution 
or a solution in which the Thomas-Fermi and WKB ones are combined.

As shown in the figures, the zeroth-order wave function $\psi_0$ does not
diverge at the turning point but the error is discernible 
around the classical turning point $x^{\rm (c)}$.
This small discrepancy is drastically improved by using 
the first-order wave function $\psi_1$. 
The error associated with the zeroth-order solution is estimated 
to be~\cite{note}
\begin{equation}
 {\rm max} \left| \frac{\psi_0 - \psi_{\rm exact}}{\psi_{\rm exact}} \right| 
\simeq 0.028 - 0.36 \epsilon  
+ {\cal O}( \epsilon^2 , \delta)~.
\label{8-3}
\end{equation}
Equation~(\ref{8-3})
implies that our zeroth-order wave functions (\ref{6-2}) and (\ref{7-6})
are suitable for $\epsilon > 0$ rather than $\epsilon < 0$;
when $\epsilon < 0$, the wave function is
unstable against a small perturbation. 

In conclusion, we proposed a divergence-free semiclassical method
that enables us to solve LSE and NLSE on an equall footing. 
Our zeroth-order solution is constructed upon a trajectry that
includes quantum corrections and therefore allows a rapidly 
converging perturbative expansion of the wave function. 
Our method may therefore be applied to drastically improving related 
semiclassical methods such as instantons~\cite{Raj} and 
periodic orbital theory~\cite{Gat}.  
The results of these subjects will be reported elsewhere. 

This work was supported by a Grant-in-Aid for Scientific Reseach 
(Grant No.~11216204) by the Ministry of Education, Science, Sports, 
and Culture of Japan, and by the Toray Science Foundation.  
 
\newpage

%%%%%%%{references}%%%%%%%%%%%%%%%%%%%%%%%%%%%%%%%%%%%%%%%%%%%%%%%%%%%%

%%%%%%%{figures}%%%%%%%%%%%%%%%%%%%%%%%%%%%%%%%%%%%%%%%%%%%%%%%%%%%%

\begin{figure}
\begin{center}
 \leavevmode \epsfxsize=13cm \epsfbox{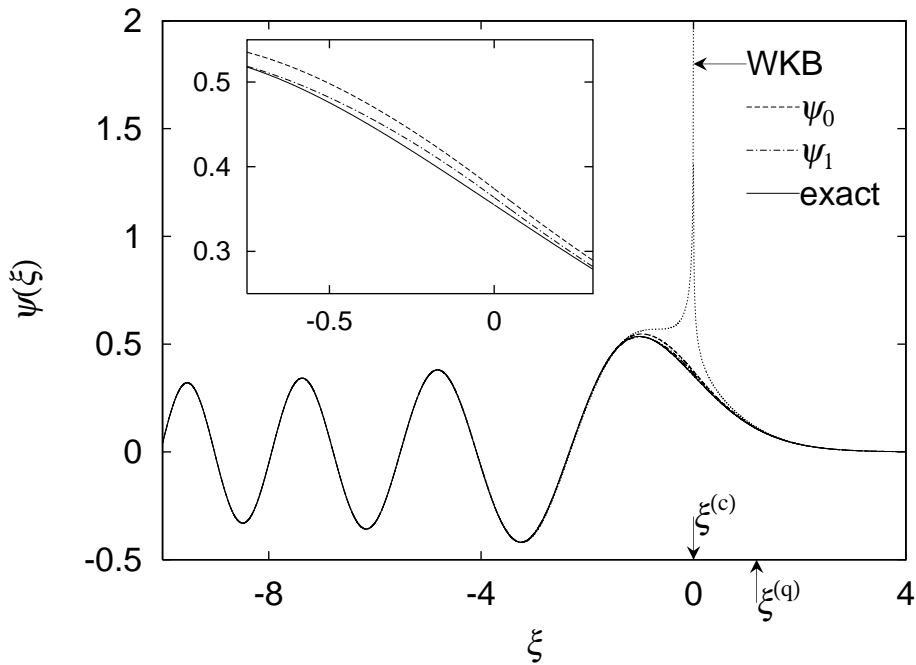}
\end{center}
\caption{Zeroth-order solution $\psi_0$ (dashed curve) 
and first-order solution $\psi_1$ (dot-dashed curve)
to LSE for a linear potential, 
$ -{d^2 \psi}/{d \xi^2} + \xi \psi = 0 $. 
The exact solution (solid curve) and 
the usual WKB solution (dotted curve)
are superimposed for comparison. 
The region around the turning point is enlarged in the inset.}
\end{figure}

%\newpage

\begin{figure}
\begin{center}
  \leavevmode \epsfxsize=13cm  \epsfbox{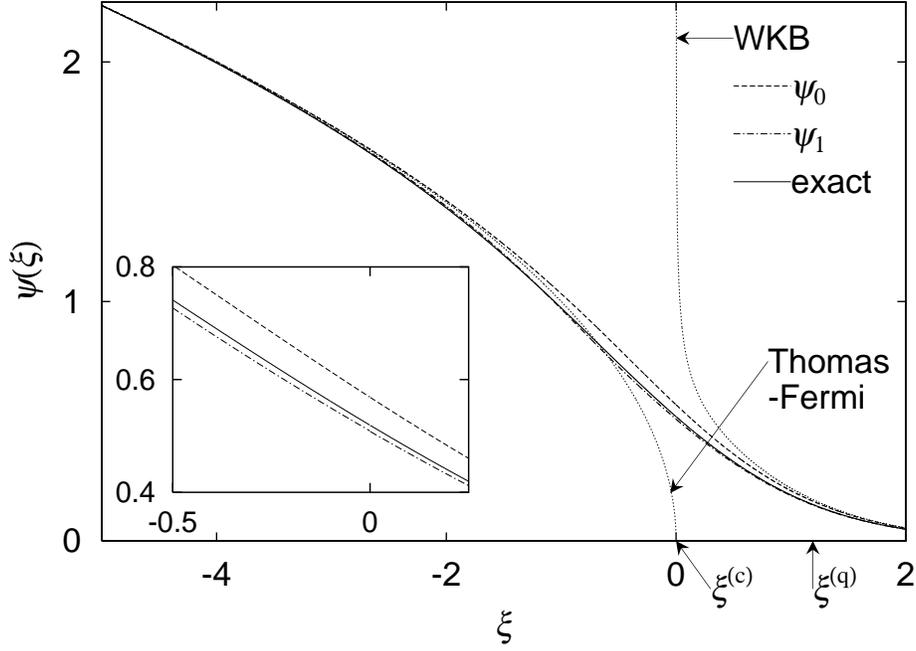}
\end{center}
\caption{Zeroth-order solution $\psi_0$ (dashed curve) 
and first-order solution $\psi_1$ (dot-dashed curve) 
to NLSE for a linear potential,
$ -{d^2 \psi}/{d \xi^2} + \xi \psi + \psi^3 = 0 $.
The exact solution (solid curve) and 
a solution in which the Thomas-Fermi and WKB ones are combined 
(dotted curve) are superimposed for comparison. 
The region around the turning point is enlarged in the inset.}
\end{figure}

%\newpage

\begin{figure}
\begin{center}
  \leavevmode \epsfxsize=13cm  \epsfbox{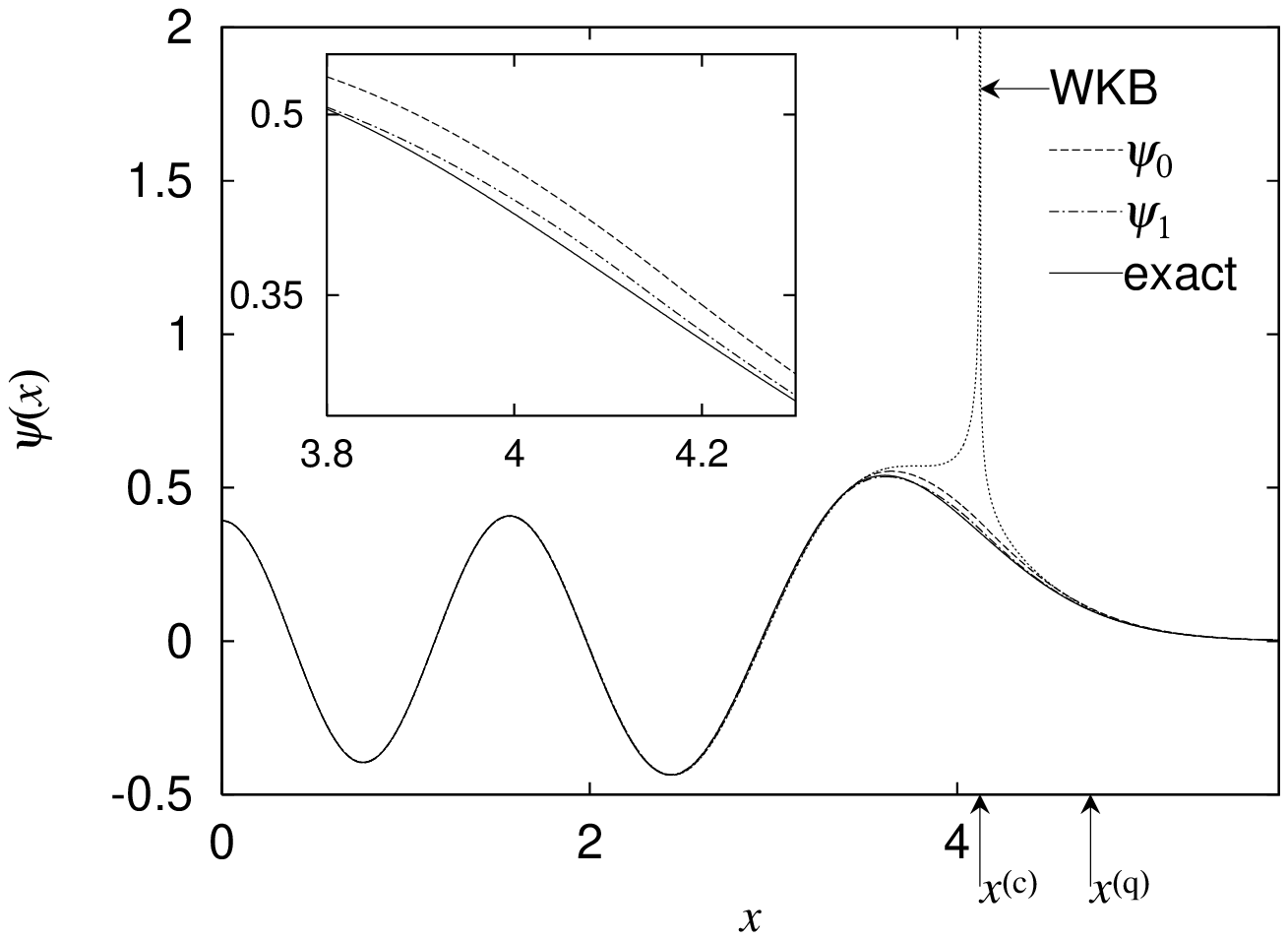}
\end{center}
\caption{Solutions to LSE for a parabolic potential,
$V=x^2$ with $E=17$ and $\epsilon=0.0476$.
The notations are the same as FIG.~1 . 
The region around the turning point is enlarged in the inset.}
\end{figure}

%\newpage

\begin{figure}
\begin{center}
  \leavevmode \epsfxsize=13cm  \epsfbox{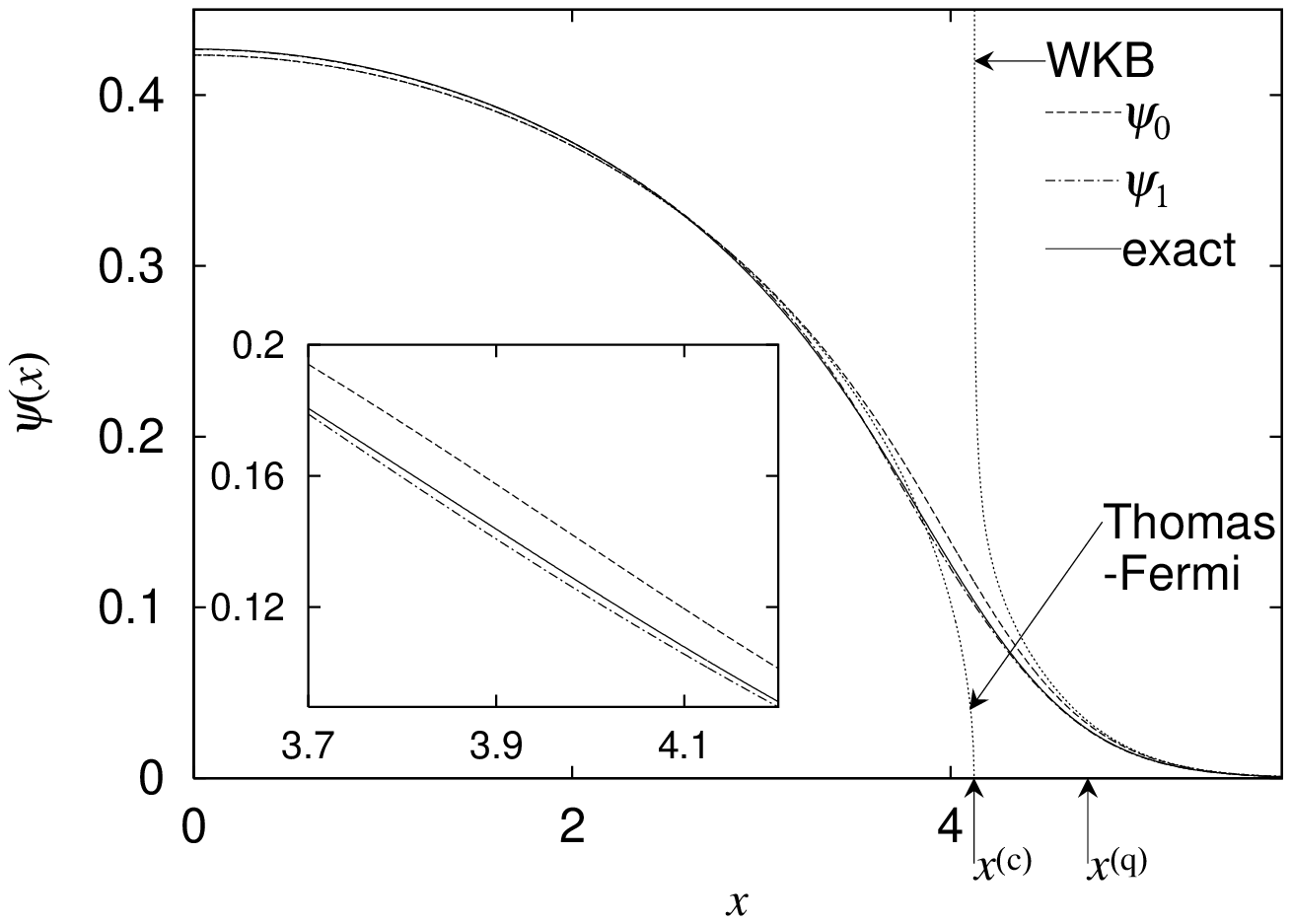}
\end{center}
\caption{Solutions to NLSE for a parabolic potential, 
$V=x^2$ with $E=17$ and $\epsilon=0.0476$.
The notations are the same as FIG.~2.
The region around the turning point is enlarged in the inset.}
\end{figure}

%\newpage

\begin{figure}
\begin{center}
  \leavevmode \epsfxsize=13cm  \epsfbox{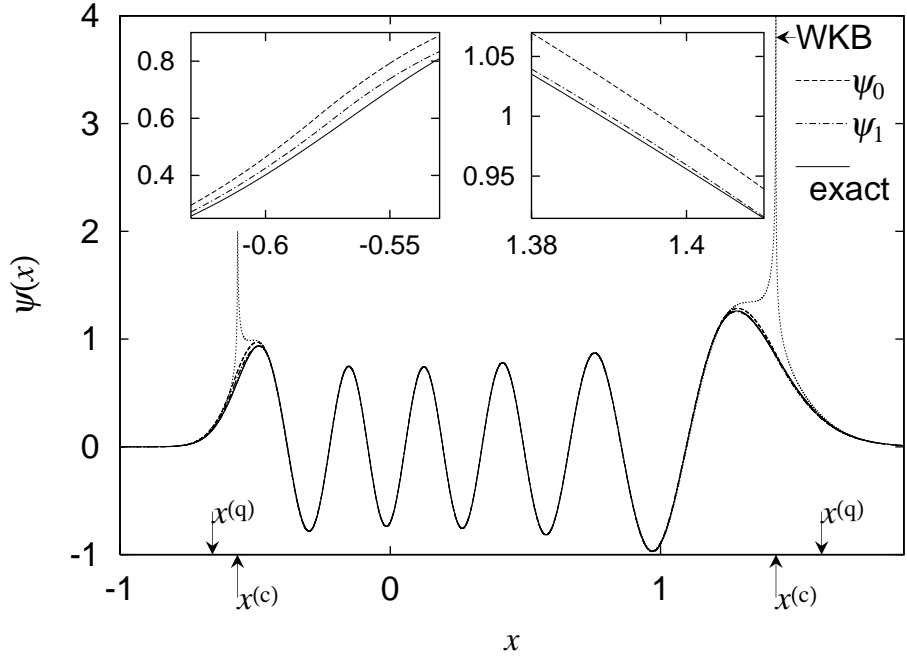}
\end{center}
\caption{Solutions to LSE for the Morse potential,
$V=900[\exp(-2x) - 2\exp(-x)]$ with 
$E=-({39}/2)^2$,
~
$\epsilon_L=0.0982$ and 
$\epsilon_R=-0.0394$.
The notations are the same as FIG.~1.
The regions around the two turning points are enlarged in the insets.}
\end{figure}

%%%%%%%%%%%%%%%%%%%%%%%%%%%%%%%END%%%%%%%%%%%%%%%%%%%%%%%%%%%%%%%%%%%%%%

\begin{references}
\bibitem{note}
Detailed derivations will be described elsewhere.
\end{references}
\end{document}